# Evolution of Complex Modular Biological Networks

Arend Hintze and Christoph Adami[*]

Keck Graduate Institute of Applied Life Sciences

535 Watson Drive, Claremont, CA 91711

[*]To whom correspondence should be addressed. Email: adami@kgi.edu

## Background

Biological networks have evolved to be highly functional within uncertain environments while remaining extremely adaptable. One of the main contributors to the robustness and evolvability of biological networks is believed to be their modularity of function, with modules defined as sets of genes that are strongly interconnected but whose function is separable from those of other modules.

## Methodology/Principal Findings

Here, we investigate the *in silico* evolution of modularity and robustness in complex artificial metabolic networks that encode an increasing amount of information about their environment while acquiring ubiquitous features of biological, social, and engineering networks, such as scale-free edge distribution, small-world property, and fault-tolerance. These networks evolve in environments that differ in their predictability, and allow us to study modularity from topological, information-theoretic, and gene-epistatic points of view using new tools that do not depend on any preconceived notion of modularity.

## Conclusions/Significance

We find that for our evolved complex networks as well as for the yeast protein-protein interaction network, synthetic lethal pairs consist mostly of redundant genes that lie close to each other and therefore within modules, while knockdown suppressor pairs are farther apart and often straddle modules, suggesting that knockdown rescue is mediated by alternative pathways or modules. The combination of network modularity tools together with genetic interaction data constitutes a powerful approach to study and dissect the role of modularity in the evolution and function of biological networks.

# Introduction

Biological function is an extremely complicated consequence of the action of a large number of different molecules that interact in many different ways. Elucidating the contribution of each molecule to a particular function would seem hopeless, had evolution not shaped the interaction of molecules in such a way that they participate in functional units, or building blocks, of the organism's function [1-4]. These building blocks can be called *modules*, whose interactions, interconnections, and fault-tolerance can be investigated from a higher-level point of view, thus allowing for a synthetic rather than analytic view of biological systems [5,6]. The recognition of modules as *discrete entities whose function is separable from those of other modules* introduces a critical level of biological organization [7] that enables *in silico* studies. Here, we evolve large metabolic networks based on an artificial chemistry of precursors and metabolites, and examine topological and information-theoretical modularity measures in the light of simulated genetic interaction experiments.

Intuitively, modularity must be a consequence of the evolutionary process, because modularity implies the possibility of change with minimal disruption of function [1], a feature that is directly selected for [3,8]. Yet, if a module is essential, its independence from other modules is irrelevant unless, when disrupted, its function can be restored either by a redundant gene or by an alternative pathway or module. Furthermore, modularity no doubt must affect the evolutionary mechanisms themselves, so that both robustness and evolvability can be optimized simultaneously [1,9,10]. A thorough analysis of these concepts requires both an understanding of what constitutes a module in biological systems, and tools to recognize modules among groups of genes. In particular, a systems view of biological function requires that we develop a *vocabulary* that not only classifies modules according to the role they play within a network of modules and motifs, but also how these modules and their interconnections are changed by evolution, i.e., how they constitute *units of evolution* targeted directly by the selection process [4].

The identification of biological modules is usually based either on functional, evolutionary, or topological criteria. For example, genes that are co-expressed and/or coregulated can be classified into modules by identifying their common transcription factor [11,12], while genes that are highly connected by edges in a network form clusters that are only weakly connected to other clusters [13]. From an evolutionary point of view, genes that are inherited together but not with others often form modules [14-16]. Yet, the concept of modularity is not at all well defined. For example, the fraction of proteins that constitutes the core of a module and that is inherited together is small [14], implying that modules are fuzzy but also flexible so that they can be rewired quickly, allowing an organism to adapt to novel circumstances [17]. Progress in our understanding of the modular nature of biological networks must come from new functional data that allow us to study different groups of genes both together and apart, and compare this data to our topological, information-theoretic, and evolutionary concepts.

A promising set of data is provided by genetic interactions [18], such as synthetic lethal pairs of genes (pairs of mutations that show no phenotype on their own but that are lethal when combined), or dosage rescue pairs, in which a knockout or mutation of a gene (in general, a loss of function) is suppressed by overexpressing another gene. Such pairs are interesting because they provide a window on cellular robustness and modularity brought about by the conditional expression of genes. Indeed, the interaction between genes—gene epistasis [19]—has been used to successfully identify modules in yeast metabolic genes [20]. However, often interacting pairs of genes lie in alternate pathways rather than cluster in functional modules, do not interact directly, and thus are expected to straddle modules more often than lie within one [21].

*In silico* evolution is a powerful tool if complex networks can be generated that share the pervasive characteristics of biological networks, such as error tolerance, small-world connectivity, and scale-free degree distribution [22]. If furthermore each node in the network represents a simulated chemical or a protein catalyzing reactions involving these molecules, then it is possible to conduct a detailed functional analysis of the network by simulating knockdown or overexpression experiments. This functional datum can then be combined with evolutionary and topological information to arrive at a more sharpened concept of modularity that can be tested *in vitro* when more genetic data become available.

Previous work on the *in silico* evolution of metabolic [23], signaling [24,25], biochemical [26,27], regulatory [28], as well as Boolean [29], electronic [30], and neural [30-32] networks has begun to reveal how network properties such as hubness, scaling, mutational robustness as well as short pathway length can emerge in a purely Darwinian setting. In particular, *in silico* experiments testing the evolution of modularity both in abstract [33] and in simulated electronic networks [30] suggest that environmental variation is key to a modular organization of function. In the experiments we describe below, we evolve large metabolic networks of many hundreds of nodes with over a thousand edges for up to 5,000 generations from simple networks with only five genes. These networks are complex—in the sense of information-rich [34,35]—are topologically interesting, and function within simulated environments with different variability that can be arbitrarily controlled. We analyze these networks using new tools that allow us to see genetically interacting pairs in the light of different concepts of modules, and compare our results to an application of those tools to the yeast protein-protein interaction network.

## Results

### Structure of the Model

**Artificial Chemistry.** We evolve the genomes of artificial cells that produce metabolites within a simple artificial chemistry of linear molecules constructed from three atoms, termed 1, 2, and 3. In valid molecules each atom must carry as many bonds as the numeral representing it, with a maximum length of twelve atoms. For example, 1-2-2-1 is a valid molecule, as is 2=2 or 1-2-3=3-2-1, but 1-3=1 is not. In this chemistry there are thus 608 valid molecules, which can undergo chemical reactions of the form A+B→A'+B' through a form of cleavage that preserves the atomic content. For example, the valid molecules 1-2-2-1 and 2=3-3=2 can react by cleaving each molecule in the middle (indicated by the arrow):

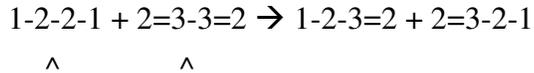

1-2-2-1 + 2=3-3=2 → 1-2-3=2 + 2=3-2-1

Of the theoretically possible cleavage reactions (cleaving any of the bonds of the 608 molecules), only 5,020,279 actually lead to valid molecules.

**Organisms.** Each organism in an evolving population consists of a cell containing molecules and proteins that perform various functions, as well as a genome (on two circular chromosomes) that codes for those proteins. The cells float in a 2D chemostat in which the smallest 53 of the 608 possible molecules are produced at a constant rate at locations from which they diffuse, and all molecules produced by the cell and exported to the environment are removed every update. The 53 short molecules play the role of precursors for the synthesis of the remaining more complex molecules. The chemostat can carry 1,000 organisms, and at each update 1 of 16 organisms is removed (see Methods).

For a cell to divide, it must produce a sufficient amount of some of the remaining 555 molecules (metabolites) within the cell, by importing any of the 53 precursors using specific transporter proteins and catalyzing any of the possible reactions with enzymatic proteins specific to the reaction. The precursors also leak into the cell at a concentration of a millionth of their concentration at the cell's location. In principle, cells can move around on the two-dimensional plane if they develop proteins for ciliates and flagella (for example, to follow the source of the precursor molecules), but these are turned off for the present experiments, so that the cells are anchored to the center of the chemostat. A description of enzyme and transporter affinities to molecules, as well as details of the calculation of organismal fitness as a function of the metabolites the cell produces is found in the Methods.

Proteins are encoded in the genome using the alphabet [0,1,2,3]. Each gene starts with four consecutive zeros (start codon), followed by the expression level, the type of protein (import, export, or catalytic), followed by the specificity to the reaction and the affinity to the molecule transported or catalyzed (see Methods). The genomes are evolved with a standard Genetic Algorithm with fitness-proportional selection (Wright-Fisher model), a Poisson-random point mutation rate $\mu=1$ per genome (but capping the maximum number of mutations per genome at six), and the possibility of gene duplication and deletion (see Methods).

**Environments.** In order to simulate dynamic and unpredictable environments, we designed three environments that differ in their precursor availability. In all environments the sources of the 53 precursor molecules are randomly distributed, and constantly replenished so that they cannot be drawn down. In the static environment, the location of the precursor sources is fixed throughout the experiment, while in the quasi-static environment the location of a single random precursor is moved each update. In the dynamic environment, the source of *all* precursors is moved every update, and 25% of the precursors are randomly chosen to be unavailable. The set of unavailable precursors also changes periodically. Most experiments were repeated in each of these environments.

**Organism and Network Evolution.** Cells are initialized with a genome encoding five genes: two proteins catalyzing molecular reactions that produce metabolites that contribute to fitness, one that produces a metabolite that does not contribute to fitness, one import protein and one export protein (see Methods). Different metabolic pathways evolve depending on the imported molecules and their abundance, and can be represented by a network connecting molecules and proteins. For example, the pathway importing molecule 1-2-1 with protein A, molecule 1-2-2-1 with protein B, and catalyzing the reaction 1-2-1 + 1-2-2-1 → 1-1 + 1-2-2-2-1 with protein C and subsequent export of 1-1 using protein D (Fig. 1A), can be represented as a graph in at least three different ways. The functional graph (Fig. 1B) uses both proteins and substrates as nodes, and connects them with edges. The metabolic graph (also called substrate graph [36], Fig. 1C) removes the proteins and places edges between substrates connected via an enzyme. In a protein-protein interaction graph all substrates are stripped, leaving only the interaction between enzymes and transporters as in Fig. 1D. The different renditions of the same pathway as networks lead to different topological properties.

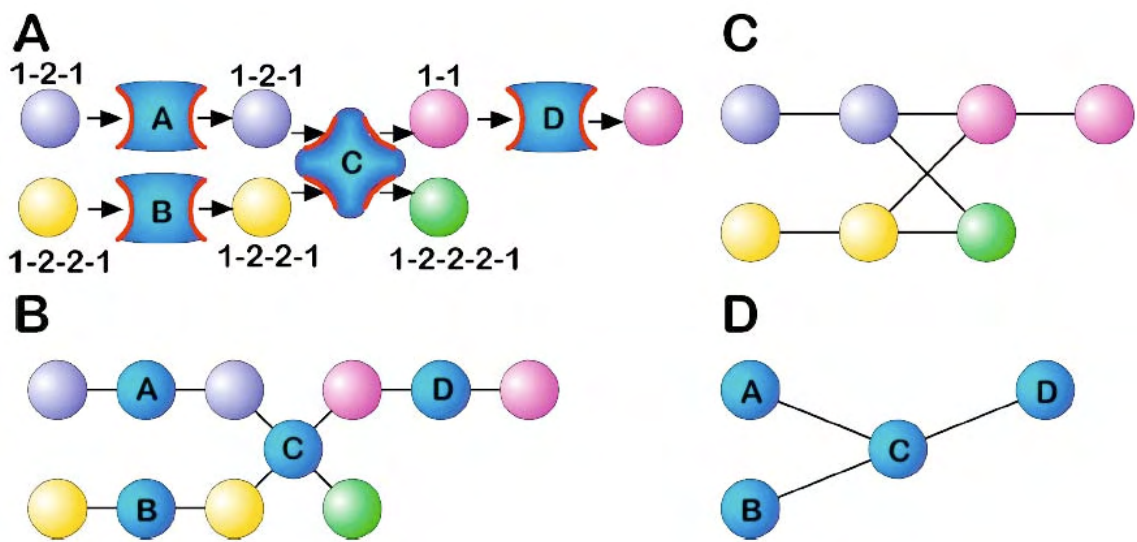

**Figure 1**. **Representations of a metabolic pathway**

Pathway importing precursors 1-2-1 and 1-2-2-1 using transport proteins A and B respectively, producing molecules 1-1 and 1-2-2-2-1 via enzymatic protein C, and exporting the by-product 1-1 using protein D. **(A):** Pathway, **(B):** functional graph, **(C)**: metabolic graph, and **(D)**: protein-protein interaction graph.

**Phylogenetic Depth.** In asexually evolving populations, every organism has a unique line of descent that connects it to the ancestral genome, via intermediary genomes carrying heritable genetic differences between mother and daughter genome that occurred during reproduction. Often these changes are single substitutions, but can also be duplications or deletions of genomic sequences of various lengths. Because the environments present the same niche to every organism, the lines of descent coalesce quickly to a single dominating type irrespective of the depth. Since beneficial mutations are very common, the phylogenetic depth is a good proxy for the number of generations elapsed in a run up to that depth.

# Network Evolution

Networks evolve to be highly complex, increase in size and develop complex pathways to metabolize the precursors. Typically, pathways evolve first via duplication and divergence of the existing genes, but later pathways are combined and new pathways emerge by evolving import proteins for precursors that leak into cells and for which catalytic proteins had evolved. Reaction networks are complicated, involving loops and multiple interconnections.

**Genetic Information Content about Environment Increases in Evolution.** In the example experiment depicted in Fig. 2, genomes evolve to close to their maximum size of 60,000 base 4 coding positions—from hereon referred to as "base pairs" (bps)—from an initial size of 2,000 base pairs (of which only 880 are functional) with an information content of approximately 36 bps or 72 bits (see Methods). In order to study the evolution of function, we followed the evolution of fitness, the number of nodes and edges of the network, and the genome's information content (as described in Methods), along the line of descent of the population. The order of a genome in the line of descent is given by the genome's phylogenetic depth from the ancestral genome (see Methods).

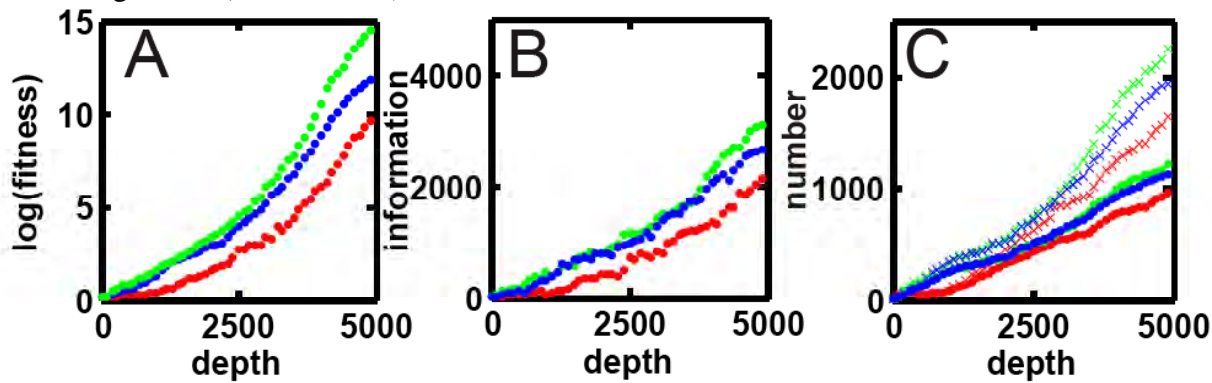

**Figure 2. Evolution of complex networks**

**(A)** Log (base 10) of fitness along the line of descent, starting with the ancestor (phylogenetic depth zero) to the $5,000^{th}$ organism on the line, for a static (green), quasi-static (blue) and a dynamic (red) environment. **(B)** Information content of genes (measured in base pairs) along the line of descent every 100 generations, colors as in (A). **(C)** Evolution of number of nodes (points) and edges (crosses) along the line of descent. Colors as in (A).

We show in Fig. 2 the fitness, information content, and number of nodes and edges for three runs in different environments, for every 100th organism on the line of descent, to a depth of 5,000. The information content increases in lock step with the fitness, indicating that the information content is a good proxy for the functional complexity of the cells. We found no evidence that the amount of information that is acquired ultimately depends on whether the environment is dynamic or not. However, networks evolve more slowly in dynamic environments because the unpredictable environment requires more complex pathways for the organism to function reliably.

**Evolved Metabolic Networks Have Pervasive Properties.** The metabolic networks generated by the evolved genomes can be analyzed using standard tools, and display some of the usual properties that distinguish biological networks from random graphs [22]. Fig. 3 shows the average degree distribution obtained from 80 networks independently evolved to depth 1,000 in a dynamic environment, and binned using a threshold binning method [37]. The distribution depends on whether a functional, metabolic, or protein-protein interaction graph (as defined in Fig. 1) is drawn. Both the functional and the metabolic network appear approximately scale-free, a finding commensurate with an analysis of the degree distribution of the central metabolic network of *E. coli* [36]. The distribution for the protein-protein interaction graph (Fig. 3C) is exponential on the other hand, and similar to that of a random modular network [38]. On the contrary, the protein-protein edge distribution found in Ref. [36] for *E. coli* is more power-law like than the one we find here, but this evidence is weak due to a long exponential tail. Note that the probability to have four edges deviates from the power law in Fig. 3A because all reactions are of the form A+B→A'+B' in this model.

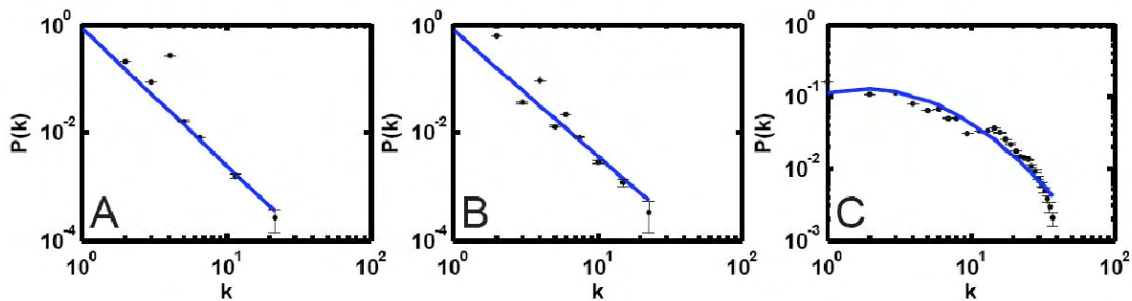

**Figure 3. Edge probability distribution for evolved networks**

Probability distribution $P(k)$ based on **(A)** functional, **(B)** metabolic, and **(C)** protein-protein representation of interactions. The functional distribution decays approximately as $P(k) \sim 1/k^\gamma$ with $\gamma = 2.53$ (minimum T=20 points per bin, $r^2$=0.91, fit not shown). Excluding the point at $k$=4 yields $\gamma = 2.52$ with $r^2$=0.97 (blue line, T=20). The metabolic distribution decays with $\gamma \approx 2.34$ (line, T=20, $r^2$=0.87). The protein-protein distribution was fitted to an exponential (binning threshold T=100, $r^2$=0.88). Error bars are standard error.

The probability distribution that a substrate participates in $k$ metabolic reactions is also a power law, with $p(k) \sim k^{-\lambda}$ with $\lambda \approx 2.23$ (Supporting Figure S1). A similar value was found empirically for this distribution in the *E. coli* metabolic network [22].

The paths between nodes in the network (the "average geodesic distances", see Methods) are short ("small-world networks"), normally distributed (Supporting Figure S2), and they remain short even as the network size grows during evolution (Supporting Figure S3). This small-world character has been shown to be a universal feature of metabolic networks in 43 organisms [22,36], and is hypothesized to be an adaptation geared towards minimizing the transition time between metabolic states when reacting to changed external conditions.

Similar to what was observed in yeast protein-protein interaction networks [21], the path length in our networks increases dramatically up to a break point when nodes that are characterized as hubs are removed from the network (see Fig. 4), but increases smoothly until the network almost collapses if random nodes are removed instead.

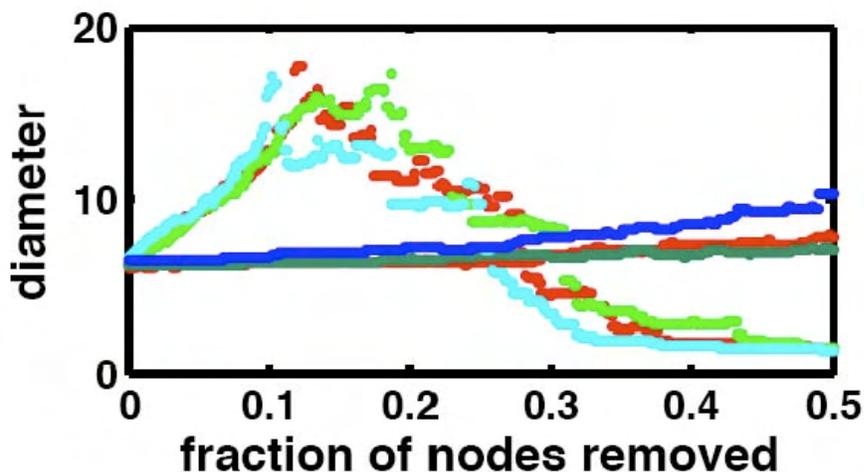

**Figure 4. Average diameter (path length) under node removal**

Average network diameter at depth 5,000 under node removal, for the functional network. Light colored dots: path length with removal of hubs, dark colored dots: path length with removal of random nodes. Green: static environment, blue: quasi-static, and red: dynamic environment. The breakdown under hub removal comes at about 200 hubs removed.

In Fig. 5, we show a network evolved in a dynamic environment, with 534 genes producing 435 molecules, with nodes representing molecules and proteins (functional network annotation).

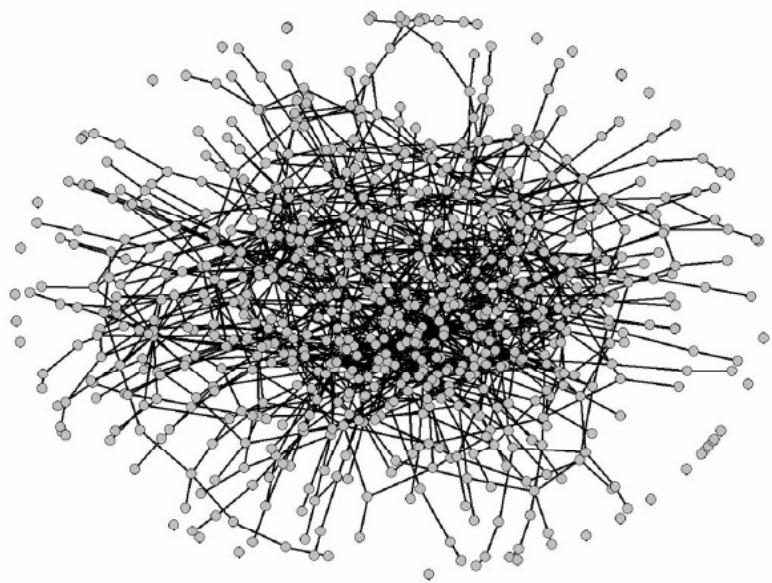

**Figure 5**. **Evolved metabolic network**

The functional network of a cell with phylogenetic depth 5,000, with 969 nodes and 1,698 edges, rendered with PAJEK [39].

**Network Modularity Increases in Evolution.** We can assign a network modularity score to every network on the evolutionary line of descent using the information bottleneck algorithm of Ziv et al. [40] as described in Methods. The modularity of the networks increases over evolutionary time in the long run, but can go up or down intermittently as new pathways are forged. Figure 6 shows an average of the modularity measured in nine independent runs performed in dynamic and static environments. While the network modularity is similar when the networks are small, the modularity score of networks evolved in dynamic environments is significantly lower on average for most of the time. To cope with the unreliable precursor supply in dynamic environments, networks that evolve in such environments ensure the presence of precursors by evolving the requisite production pathways and integrating them into the metabolic pathways. The precursor reactions effectively *connect* the main metabolic pathways. Indeed, the fraction of genes involved in the production of precursor molecules increases dramatically for networks evolving in dynamic environments (Supplementary Figure S8), and only decays due to the increased production of metabolites later, when the presence of precursors can be relied upon. In other words, while functional pathways emerge both in the static and the dynamic networks, these pathways are connected by precursor reactions (and as a consequence overlap) for the dynamic environments, whereas they can remain separate for networks evolved in static environments.

Our finding that networks evolved in dynamic environments are lass modular than those evolved in static environments appears to run opposite to the conclusion reached by Kashtan and Alon [30], who noted that dynamic environments are necessary for the evolution of modularity. However, metabolic networks are very different from the type of logical networks evolved there, as is the nature of environmental changes. Our dynamic environments change randomly, whereas Kashtan and Alon's environment changes in a modular fashion, rewarding one or the other function in turn. We further comment on this observation in the Discussion.

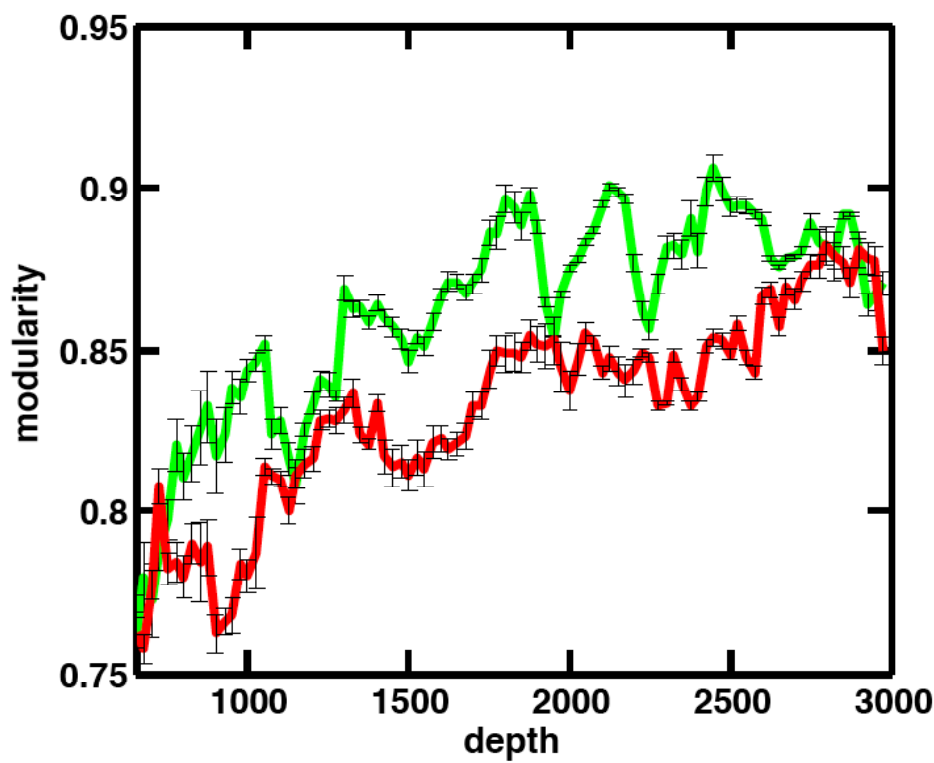

**Figure 6. Network modularity as a function of evolutionary depth**

Network modularity score for networks evolving in a dynamic (red line) compared to a static environment (green line). Each line represents the average over nine independent runs. Errors are standard error. Because the modularity score can only be calculated for networks of sufficient size, we show the modularity starting at a depth of 650.

**Mutational and Environmental Robustness Decrease.** Biological networks have evolved to be robust to mutations, knockouts, and environmental noise, as compared to random networks [3]. This robustness is believed to be due to genetic redundancy [41] as well as to the interaction between unrelated genes that can compensate for loss of function [42]. We have measured the robustness of our evolved networks to node removal as well as to environmental noise, by measuring the fitness of cells as more and more nodes are removed, and as more and more of precursor molecule concentrations are set to zero. The scaled fitness of cells decreases approximately exponentially with the number of nodes or precursors removed (see Supporting Figures S4A,B), with a fitness decay parameter that reflects the fitness effect of accumulating mutations (see Methods). The larger this parameter the more fragile the organism; consequently we define robustness as one minus fragility. We show the robustness parameter $\rho_{KO}$ and $\rho_{ENV}$ along the line of descent in Fig. 7. Node removal robustness ($\rho_{KO}$) barely decreases as the networks become more fit (even though the fitness effect is scaled to the wild-type fitness), independently of the type of environment. Environmental robustness ($\rho_{ENV}$) decreases for the static and quasi-static environments, but remains nearly constant for the dynamic environment.

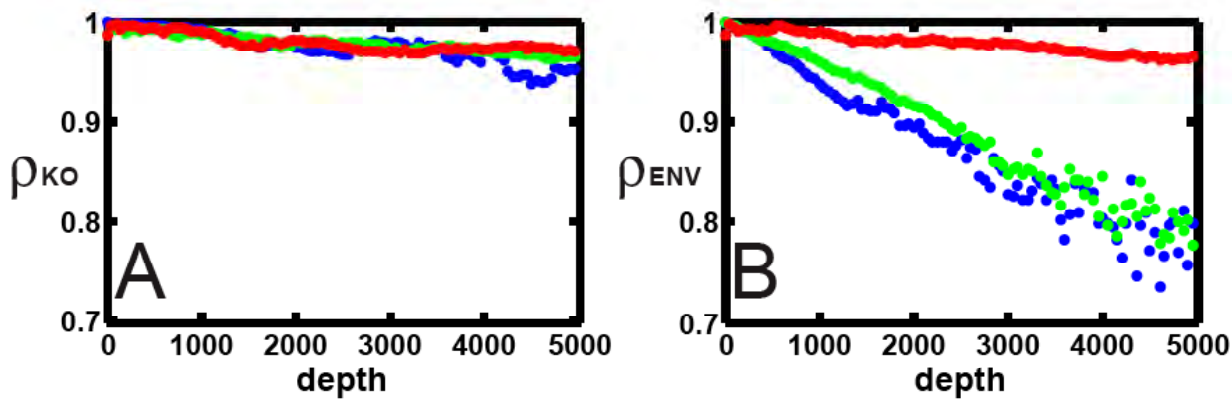

**Figure 7. Evolution of robustness**

(**A**) Node removal ($\rho_{KO}$) and (**B**) environmental ($\rho_{ENV}$) robustness along the line of descent for a static (green), quasi-static (blue) and a dynamic (red dots) environment as a function of phylogenetic depth.

**Genetic Interactions and Modularity.** To understand how modules interact, we studied whether genetic interactions occur predominantly between genes within modules or between modules, for the networks evolved in dynamic vs. static environments. We used two different methods to determine clusters: a topological one (betweenness-centrality clustering), and an information-theoretic one (network bottleneck method, see Methods). For both of these methods, the clustering method returns a ranked list of nodes, but the orders are different, and they reflect different properties of the nodes. Modules are often thought to communicate with each other via nodes with high *betweenness centrality* (BC) [43]. Such nodes are distinguished not by their connectivity, but by being major signal thoroughfares: the shortest path of many pairs of nodes runs through them ([44,45], see Methods). To test the modular structure of our networks, we remove nodes with high BC one by one in the order of their (reiterated) BC rank, and study the rate at which pairs of nodes with a given character are separated, i.e., the shortest path between them is severed.

We obtained a list of synthetic lethal pairs by finding all those pairs of genes whose knockout does not affect fitness on their own, but cause a loss of fitness when knocked out together. Such pairs (for the network shown in Fig. 5 we found 44 of them) tend to stay together (red line in Fig. 8A), suggesting that synthetic lethals tend to cluster together within modules, and are only weakly affected by the removal of nodes with high BC. This is reflected in the distance distribution: synthetic lethals tend to be very close to each other. We also studied genes that interact via knockdown suppression or rescue. A gene rescues the knockdown (technically, downregulation by a factor 10) of another gene if the overexpression (upregulation by a factor of two) of that gene restores—even partially—the loss of metabolites suffered by the cell upon knockdown of the other gene. We ranked the pairs of genes by the absolute amount of recovered loss, that is, pairs where a knockdown only led to a small loss of metabolites are ranked lower, even if all of that loss was recovered by overexpression of the partner gene. For the analysis in Fig. 8A, we used the top 95% of all knockdown suppression pairs (7,290 pairs). Using different cutoffs (see Supporting Figure S6) does not change the picture appreciably.

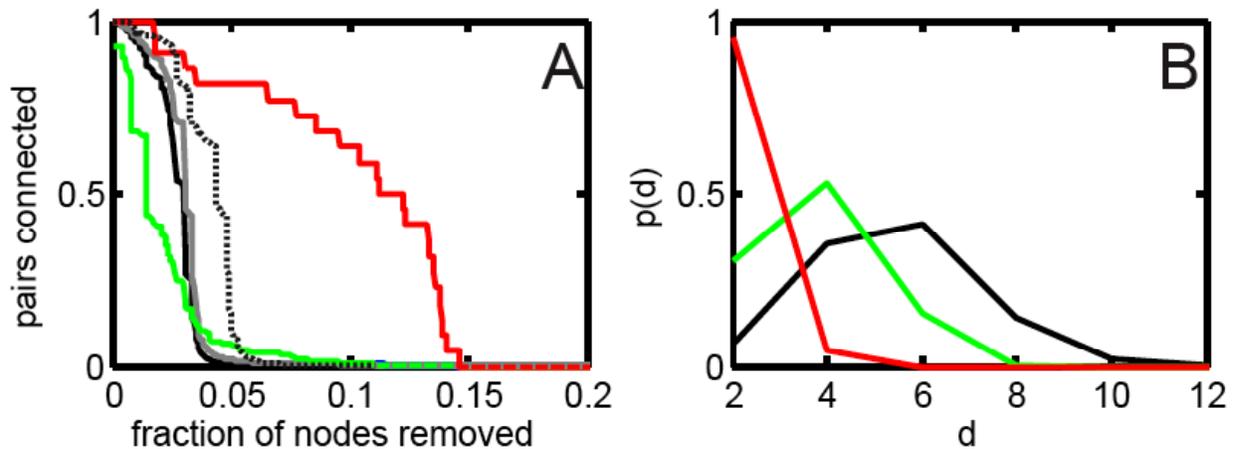

**Figure** 8. **Modularity analysis and distance distribution for topological clustering**

**(A)** Fraction of pairs of genes that remain connected upon removal of nodes with the highest betweenness centrality, for the evolved network depicted in Fig. 5. Red line: synthetic lethal pairs, green: knockdown rescue pairs, black solid line: random pairs, black dotted line: random pairs of a random network, grey line: relative size of largest connected component. **(B)** Distance distribution of pairs of genes. Colors as in **(A)**.

The rate at which pairs of genes are separated is explained in part by their distance distribution (Fig. 8B): the closer two genes are in a network, the lower the probability they are broken up by removing nodes with high BC. Note that we have omitted odd distances in Fig. 8B because in the functional network these are represented by molecules, whereas the pairs studied here are proteins. Thus, the distance between any two proteins is even. As a baseline for comparison, we use the rate at which random pairs of genes are separated (black line in Fig. 8A).

When removing nodes with high BC, knockdown suppressor pairs (green) are separated quickly, in fact much more quickly than is suggested by their distance distribution, which peaks in between that of the random pairs and the synthetic lethal pairs (Fig. 8B). A neutral assumption would be that distant pairs have a higher chance to be disrupted by a node removal. Instead, random pairs (whose average distance is the largest) stay connected much longer than knockdown suppressor (or more generally, compensatory) gene pairs. This is possible if compensatory gene pairs are preferentially connected by nodes with high BC, or are themselves nodes with high BC. Since nodes with high BC are thought to connect modules, we can deduce that compensatory gene pairs preferentially straddle modules.

We also studied how the decay of genetically interacting pairs compares to global topological properties, and compared their behavior to similar experiments performed in random networks. The size of the largest connected component in the functional network (grey line in Fig. 8A) decays somewhat more slowly than random pairs because disrupting such pairs does not necessarily change the connected component. We can also ask whether the peculiar scale-free degree distribution is dictating the behavior of the random pairs under node removal. The fraction of random pairs of nodes from a randomized network with the same number of nodes, edges and degree distribution as our evolved network (black dotted line in Fig. 8A) is decreasing much more slowly, however, indicating that they are not separated by nodes with high betweenness centrality. In other words, random networks—even when constructed to have precisely the same degree distribution as our functional networks—are not modular in a topological sense. Very similar conclusions can be drawn from networks evolved in static or quasi-static environments (see Supplementary Figure S5).

We can compare the behavior of these genetically interacting pairs in evolved metabolic networks to equivalent pairs in the highly-curated yeast protein-protein interaction network of Reguly et al. [18] of 1,038 nodes with the genetic interactions removed. For this network, both synthetic lethal and compensatory pairs are separated *later* than random pairs (list of synthetic lethal and dosage rescue pairs from Reguly et al. [18]). However, synthetic lethals are still separated later than compensatory pairs of genes, suggesting that synthetic lethals preferentially occur within rather than between, modules. Interestingly, pairs of nodes from a randomized yeast network (retaining number of nodes, edges, and edge distribution) decay more slowly than compensatory genes, just as in our metabolic networks. We discuss the difference between our evolved metabolic networks and the yeast protein-protein interaction network as revealed by this analysis further below.

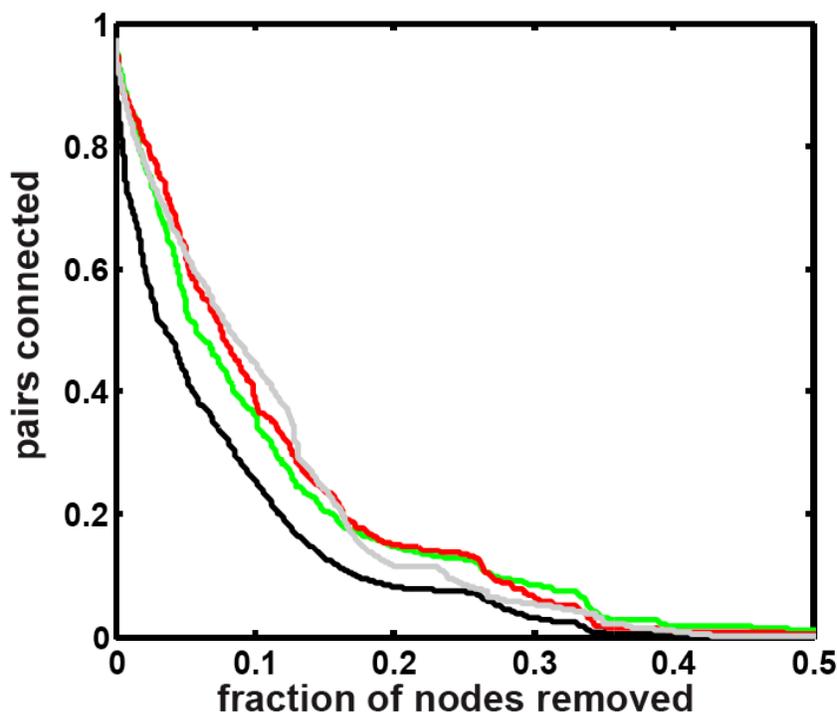

**Figure 9. Modularity analysis for the yeast protein-protein interaction network with topological clustering**

Fraction of pairs of genes that remain connected upon removal of nodes with the highest betweenness centrality, for the highly curated yeast protein-protein interaction network and synthetic lethal and dosage rescue pairs of Reguly et al. [18]. Red line: synthetic lethal pairs, green: dosage rescue pairs, black solid line: random pairs, grey line: random pairs of the randomized network.

We can also study the relationship of genetically interacting pairs with clusters determined by an information-theoretic method [40]. Clusters determined by this method are chosen so that they simplify the original network while the relevant character of the network (the fidelity) is determined by network diffusion (see Methods). This algorithm results in a list of nodes that reflects the order in which nodes are merged to generate the optimal clustered network. We can use this list to study the fraction of genetically interacting pairs that remain separate under the node merging procedure (shown in Fig. 10) in a manner similar to the procedure used to obtain Fig. 8. However, the order of the nodes in the information-clustering list is fundamentally different from that reflecting betweenness centrality: the nodes that are merged first into modules are by definition those that are information-theoretically redundant and are close to each other under graph diffusion. Thus, we expect the pairs that are close to each other in graph diffusion distance to be merged first. Fig. 10A shows the rate at which synthetic lethal and knockdown suppressor pairs remain separate for the largest connected component of the same network as analyzed in Fig. 8A. The same analysis applied to the largest connected component of the yeast protein-protein interaction is shown in Fig. 10B, where dosage rescue pairs are used as a surrogate for knockdown suppressor pairs. In both cases, random pairs remain separate the longest while genetically interacting pairs are merged much earlier. In yeast, synthetic lethals are merged earlier than dosage rescue pairs, suggesting again that synthetic lethals are preferentially found within modules. The situation is less clear for our artificial metabolic networks. There, synthetic lethals and compensatory (suppressor) pairs are separated at different rates, depending on the coarse-graining of the network. Still, genetically interacting pairs are markedly different from random pairs of genes under this procedure, highlighting their importance for the study of modularity.

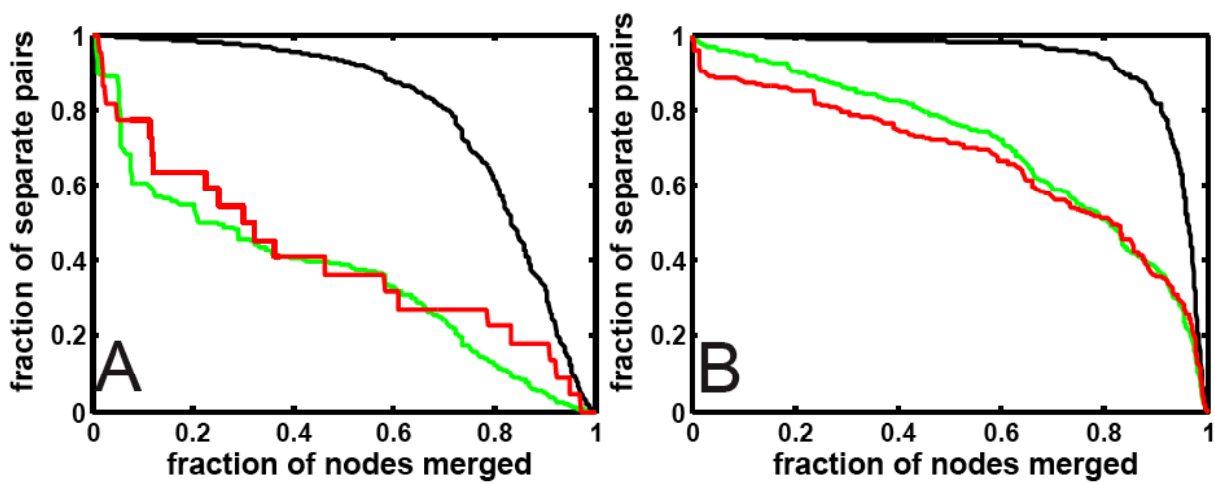

**Figure 10. Modularity analysis using information-theoretic clustering**

Fraction of pairs remaining separate as nodes are merged according to the information bottleneck clustering. (**A**) Largest connected component of a metabolic network evolved in a dynamic environment. Red line: synthetic lethal pairs, green line: knockdown suppressor pairs, black: random pairs. (**B**) Largest connected component of the yeast protein-protein interaction network. Colors as in (A), but compensatory pairs are dosage rescue pairs.

## Discussion

Evolution shapes our artificial metabolic networks into complex tightly connected pathways that are modular in nature, and that share many of the well-known properties of biological networks, such as scale-free edge distribution, small-world connectivity, and hubness. We can use these networks to study how established concepts of modularity—such as betweenness centrality clustering and information-theoretic modularity—compare to the rate at which genetically interacting pairs are disrupted by either removing nodes with high BC, or merging nodes that have been assigned to the same information-theoretical cluster. By evolving networks in different environments that are expected to yield different modularities, we can dissect the impact of genetically interacting pairs on modularity notions. When we compare the behavior of genetically interacting pairs in our evolved networks to those in the yeast protein-protein interaction network, we find commonalities and some discrepancies.

One of our main finding is that synthetic lethal pairs usually lie within modules, no matter how modules are defined, and that compensatory (suppressor) pairs preferentially straddle modules. We also find that in our metabolic networks, many nodes that are assigned the same module in fact have high betweenness centrality themselves, a property that does not appear to be shared with the yeast protein-protein interaction graph, where random pairs separate faster than compensatory pairs. A number of differences between the networks can explain these findings. First, the functional graphs (Fig. 2B) we use to determine nodes of a network have a different connectivity pattern than protein-protein interaction networks as shown in Fig. 3, and are sparser. Second, the multi-copy suppressor pairs we use to mark genetic compensation in our metabolic networks are different in nature from the dosage rescue pairs listed in Reguly et al. [18]. Also, synthetic lethality for metabolic networks refers almost exclusively to functional redundancy, whereas synthetic lethality in yeast can involve complex and indirect interactions. While in principle we could have restricted the comparison of our evolved networks to only the metabolic component of the yeast interaction network, the number of genetically interacting pairs of genes affecting metabolic genes in Reguly et al. is not sufficient to establish significance. Experimental work in progress by several groups to obtain a large number of multi-copy suppressor pairs in yeast will change this situation dramatically.

We find no evidence that dynamic environments are required for the evolution of functional modules [30,33]. Rather, it appears that genes segregate into functional modules as long as there are a large number of different ways to achieve functionality. Indeed, on the contrary, metabolic networks evolved in dynamic environments appear to be less modular. We can understand this finding by noting that our dynamic environments change randomly by omitting the availability of a random fraction of precursors, as opposed to the modular changes implemented in Ref. [30]. To deal with the unpredictability of the environment, our metabolic networks first evolve reactions that produce precursors from other precursors and metabolites (see Supplementary Figure S8) such that several different genes produce the same precursor from different precursors and metabolites at any point in time. In that way, the evolved redundancy ensures the presence of any particular precursor. Because this redundancy creates connections between pathways, the modularity score of such networks is lower. We also find that networks evolve more slowly in dynamic environments, but they are more robust to environmental fluctuations in return. Thus, at least for metabolic networks, robustness and modularity do not necessarily go hand-in-hand.

The *in silico* evolution of functional networks based on artificial genetics and chemistry presents an opportunity to study how complex networks, their structure and organization, evolve over time to cope with environments with varying degrees of predictability. We believe that such networks can provide a formidable benchmark for experiments with biochemical networks, and allow predictions with hitherto unavailable accuracy. The type of functional interaction experiments that we performed on our large evolved networks anticipates high-throughput efforts currently under way using temperature-sensitive yeast deletion mutants and their multi-copy suppressors, and suggests that dosage rescue (or multi-copy suppressor) pairs of genes represent an appropriate and sensitive tool to study modularity in biological networks.

## Methods

**Genome Code and Organization.** Molecular interactions occur through proteins that catalyze the reactions between the molecules of our artificial chemistry and transport them in and out of cells. These proteins are encoded by an artificial genetics using the four "nucleotides" 0,1,2, and 3 and determine the rate at which the reactions proceed. An open reading frame on a chromosome starts with four zeros (see Supporting Table 1), followed by a code indicating the expression level, followed by a tag designating the protein type, followed by the specificity and the affinity. The specificity is a 12 nucleotide stretch that determines the target molecule or reaction (e.g., if the tag is "import", 123210000000 specifies that molecule 1-2-3=2-1 is transported into the cell). Reactions are specified by mapping the 5,020,279 legal reactions to the $4^{12}$ = 16,777,216 possible 12-mer specificities, in such a manner that any mutation in the specificity region is guaranteed to catalyze a legal reaction.

A protein's affinity is determined by an "active site" that has four domains; one each for the four molecules involved in the reaction A+B→A'+B'. The binding affinity of a transport protein to the specified target is obtained by averaging the affinity of all four domains. Each domain has twelve entries that are matched to particular molecules (of maximally twelve atoms) in the following manner. First, a molecule is translated into its binary equivalent, for example, 1-2-3=2-1 is 01-10-11-10-01-00-00-00-00-00-00-00-0000-00-00 (zeros are used to pad molecules smaller than 12 atoms). The 24 bit domain of the protein $P$ is compared with the binary equivalent of the target molecule $M$, resulting in an affinity score $D(M,P)$ that is highest if the protein domain is precisely complementary to the molecule. So, for example the perfect domain for molecule 1-2-3=2-1 is 10-01-00-01-10-11-11-11-11-11-11-11. Numerically, $D(M,P)$ is obtained as $1-S(M,P)$, where $S(M,P)$ is a *similarity* score

$$S(M,P) = \sqrt{\frac{1}{108}\sum_{i=1}^{12} f^2(m_i \otimes p_i)}$$

where $f(m_i \otimes p_i)$ is the base-10 translation of the logical bitwise EQUAL of the molecule's and protein's $i$th site. The base-10 translation of the equivalent of a perfect match ('11') is 3, so that the maximal $\sum_{i=1}^{12} f^2(m_i \otimes p_i)$ is 12 x $3^2$=108, ensuring that $0 \leq A(M,P) \leq 1$. The complementarity scheme is chosen to minimize the occurrence of domains of the type 00-00-00-00, as they would be decoded as start codons. The maximal genome size in this model is 120,000 bits, or 60,000 nucleotides, on 2 circular chromosomes. Genes are allowed to overlap. Note that because of the absence of recombination, one of the two chromosomes consistently degenerates during evolution so that all of the complexity ends up contained in a single circular genome.

**Chemostat Physics and Reaction Kinetics.** Cells live in a two-dimensional space within which precursor molecules are produced at defined locations and diffuse out, so that the concentration of molecule $M$ at distance $d$ from the source, $[M](d)$, depends on the concentration at the source via

$$[M](d) = [M](0)\frac{1}{\sqrt{2\pi}}e^{-d^2/2}, \qquad (1)$$

which is the solution of the diffusion equation with a diffusion coefficient $D=1/2$, at time $t=1$.

Molecule concentrations $[M_i]$ are updated according to a discretized version of the standard metabolic rate equations [46]

$$\Delta[M_i] = \sum_{j=1}^{r} c_{ij} v_j, \qquad (2)$$

for molecules $i=0\ldots607$, where the sum runs over reactions $j=1$ to $r$, and the matrix $c_{ij}$ is the connectivity matrix of the network defined as

$$c_{ij} = \begin{cases} -1 & \text{if molecule } i \text{ enters reaction } j \\ +1 & \text{if molecule } i \text{ exits reaction } j \\ 0 & \text{otherwise} \end{cases}$$

and $v_j$ is the metabolic flux

$$v_j = \sum_{l,m} \frac{[M_l]}{k_l^{\text{out}}} R_{lm}^{(j)} \frac{[M_m]}{k_m^{\text{out}}} A^{(j)}[P_j]. \qquad (3)$$

In Eq. (3), $k_l^{\text{out}}$ is the number of edges leaving molecule $l$, and we defined the reaction matrix for reaction $j$

$$R_{lm}^{(j)} = \begin{cases} 1 & \text{if reaction } j \text{ takes molecules } l \text{ and } m \text{ as input} \\ 0 & \text{otherwise} \end{cases},$$

as well as the affinity $A^{(j)}$ by

$$A(j) = \frac{1}{4}\sum_{p=1}^{4} D(M_p, P_p), \qquad (4)$$

where $D(M_p, P_p)$ are the affinities of protein domain $P_p$ to the molecules $M_p$ as defined above.

**Organism Fitness.** The fitness of an organism is determined by the amount and complexity of the molecules it can metabolize from the precursors. The 608 possible molecules of the artificial chemistry are numbered according to their complexity (length and type of atoms):

$$M_0 : 1-1$$
$$M_1 : 2 = 2$$
$$M_2 : 3 \equiv 3$$
$$M_3 : 1-2-1$$
$$M_4 : 1-3 = 2$$
$$M_5 : 2 = 3-1$$
$$\vdots$$
$$M_{607} : 2 = 3-3 = 3-3 = 3-3 = 3-3 = 3-3 = 2 \quad ,$$

and the first 53 molecules are arbitrarily termed precursors. The remaining 555 molecules are metabolites of increasing complexity (the most complex one being $M_{607}$). Each different molecule metabolized by the cell contributes to the total fitness. If $\Delta(M_i)$ is the total amount of molecule $i$ synthesized by the cell, the total fitness is calculated using the fitness value of each the molecules $M_i$, which depends on its index $i$ via

$$\phi(M_i) = \begin{cases} 0 & (i < 53) \\ \dfrac{i^2}{608^2} & (i \geq 53) \end{cases}, \qquad (5)$$

as

$$w = \prod_{i\,\text{produced}} (1.1 + \phi(M_i)\Delta(M_i)). \qquad (6)$$

In Eq. (6), the product extends only across metabolites that have achieved non-vanishing abundance during a cell's lifetime.

Because of the explicit dependence of a cell's fitness on the concentration of precursors in the cell's vicinity, fitness is context dependent, and in principle depends on the frequency of other cells in a population. Due to the multiplicative nature of the fitness function, the discovery of new pathways is always beneficial with the same percentage, and the fitness increases exponentially during evolution. We usually plot the logarithm of the fitness, which is additive.

**Evolution.** A Genetic Algorithm [47] is used to evolve circular genomes encoding genes using the nucleotide alphabet [0,1,2,3]. Mutations are Poisson-random with a mean of one mutation per genome (and a maximum of six mutations per genome). With a probability of 1/16 per genome, a stretch of 4-512 base pairs is duplicated and inserted directly adjacent to the duplicated stretch. With the same probability, a stretch of the same size is deleted from the genome. No recombination takes place between genomes. The probability for a genome to be replicated is proportional to the fitness calculated in Eq. (6) (Wright-Fisher selection). Organisms must be at least 8 updates old before they can replicate, and they are protected from death during those first 8 updates.

**Ancestral Genome.** We designed the ancestral genome to have 3 genes on the first 1,000 bp chromosome, with the 2$^{nd}$ chromosome of 1,000 bps filled with poly-'3's in order to be as distant as possible to start codons. However, it turned out that the third gene has a start codon (0000) within its specificity domain as well as in the sequence specifying the expression level, both of which give rise to two additional proteins in overlapping reading frames. Those proteins, because they are useless to the organism, quickly disappear within the first tens of generations. The spaces between the first three genes are filled with random sequence, and the 880 bp genome is padded with 120 poly-'3's, to make up the 1,000 bp of the ancestral genome as sketched below.

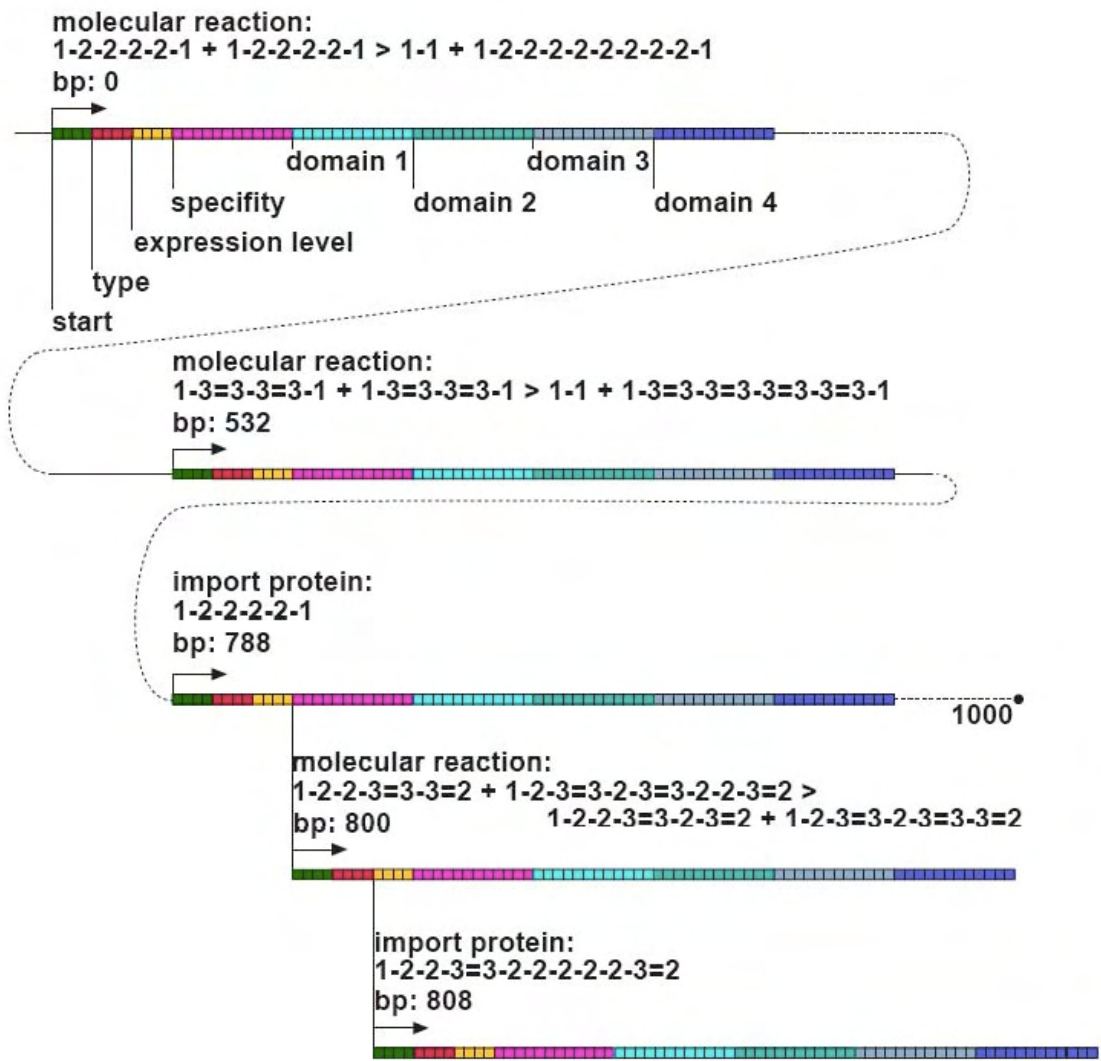

**Figure 11. Structure of the 1,000 bp ancestral genome used to start all evolutionary runs**

Each gene begins with a start codon (green), followed by type, expression level, and specificity determining regions (red, yellow, pink respectively), followed by domains encoding protein affinity. The last two reading frames (at 800bp and 808bp) are overlapping genes.

**Information Content.** The complexity of an organism can be estimated by the amount of information its genome encodes about the environment within which it thrives [34,35,48]. We can estimate the information content $I$ of a sequence $s$ of length $L$ encoding the bases 0,1,2,3 by $I = L - H(s)$, where the entropy of the sequence $H(s)$ is approximated by the sum of the per-site entropies $H(s) \approx \sum_{x=1}^{L} H(x)$, with a per-site entropy

$$H(x) = -\sum_{i=0}^{3} p_i \log_4 p_i . \qquad (7)$$

In Eq. (7), the $p_i$ are the probabilities to find base $i$ at position $x$, which can be obtained from an alignment of genomes in mutation-selection balance. For small populations and large genomes, this balance is not achieved, and the substitution probabilities $p_i$ must be estimated using the fitness effect of each substitution $w_i$ according to the implicit equation [49]

$$p_i = \frac{p_i w_i}{\overline{w}}(1-\mu) + \frac{\mu}{4}\sum_{j=0}^{3}\frac{p_j w_j}{\overline{w}}, \qquad (8)$$

where $\overline{w} = \sum_{i=0}^{3} p_i w_i$ is the mean fitness of the possible alleles at that position and $\mu$ is the mutation rate per site. We obtain the fitness $w_i$ of each allele at each position by constructing the genotype and evaluating the fitness of the cell it gives rise to in the appropriate environment. (Mutations that appear to be beneficial are counted as wild-type fitness.) Using the four values $w_i$, the probabilities $p_i$ can be obtained by iterating Eq. (8) 10,000 times or until the variance of all $p_i$ drops below $10^{-12}$.

**Information-theoretic Clustering.** To assign a modularity score to our networks, we use the information bottleneck method [50], as applied to biological networks by Ziv et al. [40]. Briefly, the method assigns clusters to the nodes of a network described by a random variable $X$ using an assignment random variable $Z$ and a relevance variable $Y$ (the bottleneck) by maximizing both the simplicity of the description (maximizing the mutual entropy between the graph and its description $I(X:Z)$) and its relevance or fidelity (maximizing $I(Y:Z)$). This is achieved via a hard clustering method that starts with a description $Z$ with one fewer nodes than $X$, then calculates the conditional probability $p(z|y)$ from a diffusion process and selects those nodes of $X$ to *merge* in the description $Z$ that result in the highest $I(Y:Z)$. This process iterates until all the nodes have been joined and the size of $Z$ is one. This procedure results in a list of nodes (from highest cluster probability to lowest) that can be used to study how synthetic lethal and knockdown suppressor pairs are merged as an alternative to the topological clustering via betweenness centrality. A modularity score for each network is obtained as the area under the information curve obtained by plotting the normalized quantities $I(Z:X)/H(X)$ and $I(Z:Y)/I(X:Y)$ against each other [40]. Because random graphs give rise to an information curve with area 0.5, any modularity score above 0.5 signals a modular organization of the network. To obtain the modularity score in Fig. 6, we averaged the modularity score of the largest, second largest, etc. connected components of the network $\mu_i$ weighted by their relative size. Thus, if the $i$th largest connected component of the network of size $N$ is $n_i$, then the average modularity score is (note that $n_i \geq 5$ is required as the modularity of smaller networks cannot be obtained)

$$\langle \mu \rangle = \sum_i \frac{n_i}{N} \mu_i . \qquad (9)$$

**Average Geodesic Distance.** The average distance $D$ of each node to any other defines the average geodesic distance of a graph

$$D = \frac{1}{m}\sum_{i=1}^{n}\sum_{j=1}^{n} d(i,j), \qquad (10)$$

where $n$ is the total number of nodes, $d(i,j)$ is the shortest path distance between $i$ and $j$, and $m$ is the total number of edges.

**Network and Environmental Robustness.** We measure the robustness of evolved networks with respect to node deletions and to changes in the precursor concentrations. Even though these perturbations are unrelated prima facie, there is evidence that mutational robustness and robustness to noise are correlated [28]. We measure mutational robustness by removing $n$ random nodes and determining the (scaled) fitness of the remaining graph $\frac{\overline{w}(n)}{w(0)}$, where $\overline{w}(n)$ is the mean of 1,000 independent fitness measurements of a network where $n$ random nodes have been removed. The fitness decreases exponentially as long as less than 30% of the nodes are removed, suggesting a ("knock-out") robustness parameter $\rho_{KO}$ defined via

$$\frac{\overline{w}(n)}{w(0)} = \exp(-n(1-\rho_{KO})) \quad . \qquad (11)$$

Environmental robustness is determined by evaluating the fitness of an organism as more and more of the 53 precursor molecules are removed. Fitness declines exponentially with the number of deleted nodes or chemicals removed, and robustness can be quantified by the slope of the decrease of log fitness, defining $\rho_{ENV}$ in a similar manner.

**Betweenness Centrality.** The betweenness centrality of a node in a network topology measures how many shortest paths go through that node. If $b_i$ is the ratio of the number of shortest paths between a pair of nodes in the network that pass through node $i$ and the total number of shortest paths between those two nodes, then the unscaled betweenness of node $i$ is $B'_i = \sum_{allpairs} b_i$, and the (scaled) betweenness centrality is [45]

$$B_i = \frac{2B'_i}{(n-1)(n-2)}, \qquad (12)$$

where $n$ is the number of nodes in the network. The betweenness centrality is positive and always less than or equal to 1 for any network.

**Software Availability.** The software to implement the artificial chemistry and genetics, as well as the evolution experiments described in this manuscript, is available at http://public.kgi.edu/~ahintze.

# Acknowledgements

We would like to thank D. Galas, H. Sauro, A. Raval, R. Rao, and N. Chaumont for discussions and critical insight, and S. Benner for discussions on artificial chemistry.

Author contributions: CA and AH conceived and designed the model, simulations, and methods. AH wrote the simulation and analysis tools, and performed the experiments and analysis. CA wrote the manuscript.

# Funding

This work was supported by the National Science Foundation's Frontiers in Integrative Biological Research grant FIBR-0527023.

# Supporting Information

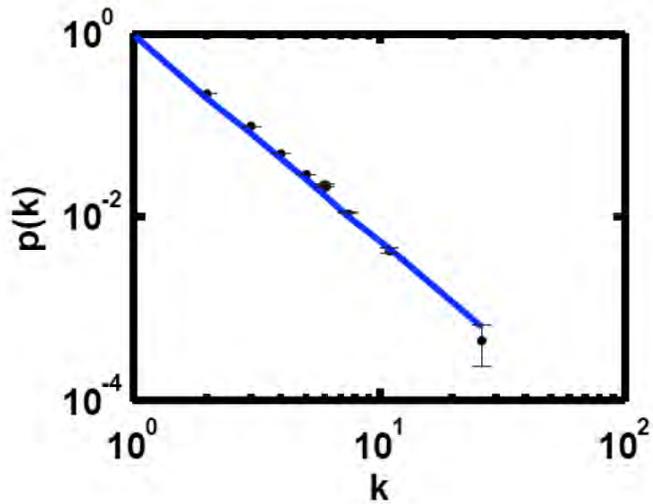

**Figure S1. Distribution of molecules in reactions**

Probability distribution $p(k)$ that a molecule participates in $k$ reactions, compiled from 80 runs to depth 1,000 in a dynamic environment. The distribution is fit to a power law $p(k) \sim k^{-\lambda}$, with $\lambda \approx 2.23$ ($r^2=0.88$). Error bars are standard error. Variable bin sizes are determined by the threshold binning method [37], with a minimum of $T=100$ points per bin.

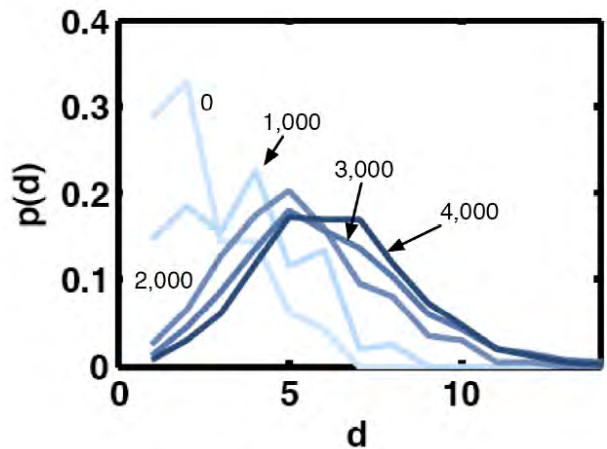

**Figure S2. Evolution of path length distribution**

Evolution of the distribution $p(d)$, the probability to find two nodes in the network that are a distance $d$ apart, for every 1,000$^{th}$ network on the line of descent, for a network evolved in a dynamic environment.

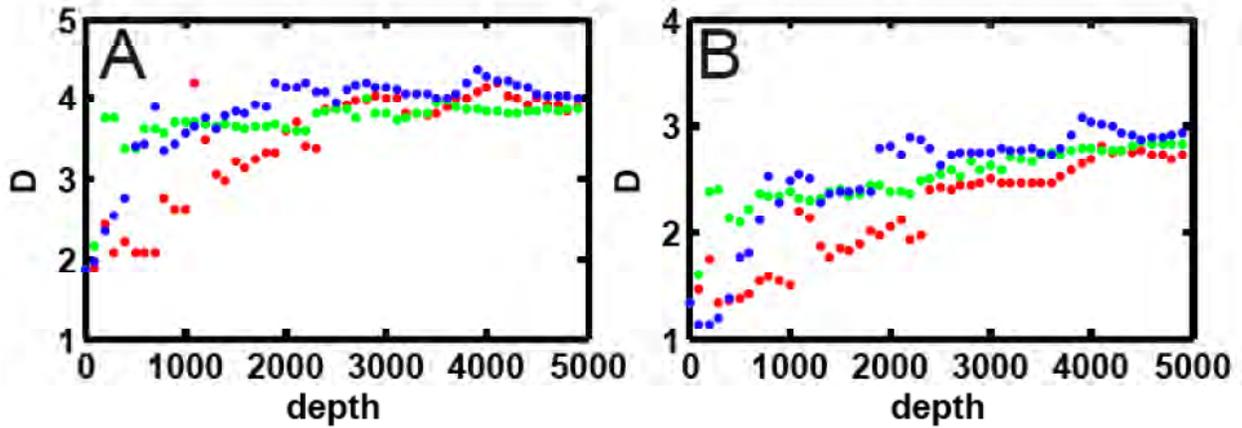

**Figure S3. Average path length D on the line of decent**

Mean path length *D* (see Methods) for a network with **(A)** metabolic, and **(B)** protein-protein annotation, in three different environments, for the network evolution shown in Fig. 2. Green: static, blue: quasi-static, and red: dynamic environment.

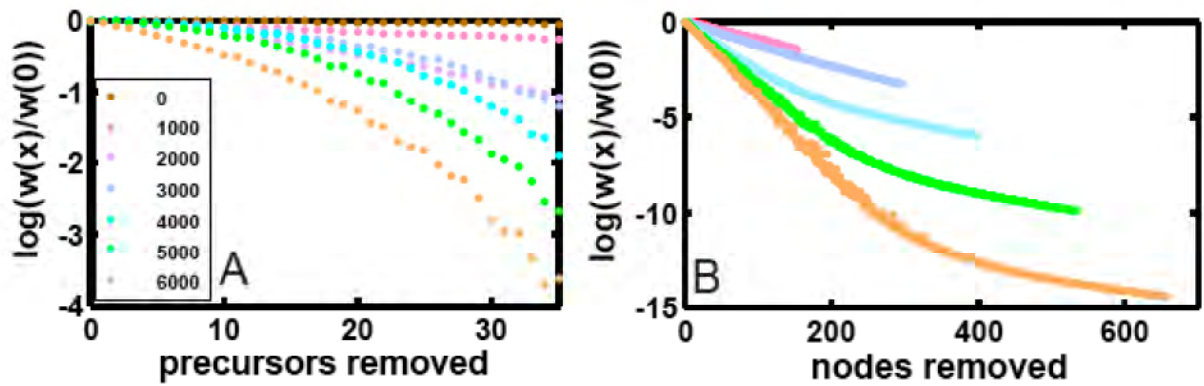

**Figure S4. Robustness of fitness under precursor and gene removal**

Decrease of normalized log fitness with increasing **(A)** precursor removal, **(B)** node removal, as a function of the position on the line of descent (colors in inset of panel A). Depth 0: ancestor.

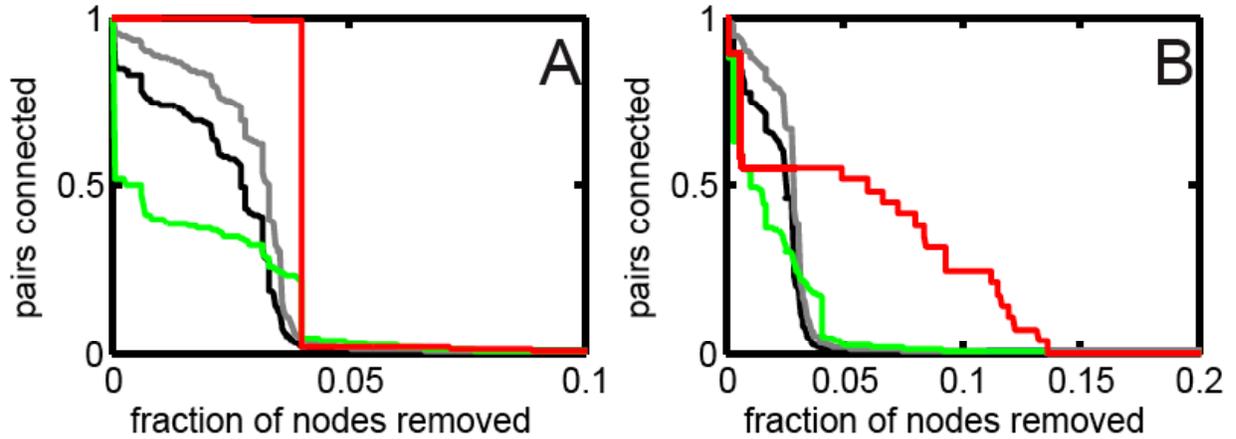

**Figure S5. Modularity analysis for static and quasi-static environments**

Analysis of the separation of pairs of genes from networks evolved in an (**A**) static and (**B**) quasi-static environment, as in Fig. 8A. Red line: synthetic lethal pairs, green: dosage rescue pairs, black: random pairs and grey: relative size of largest connected component.

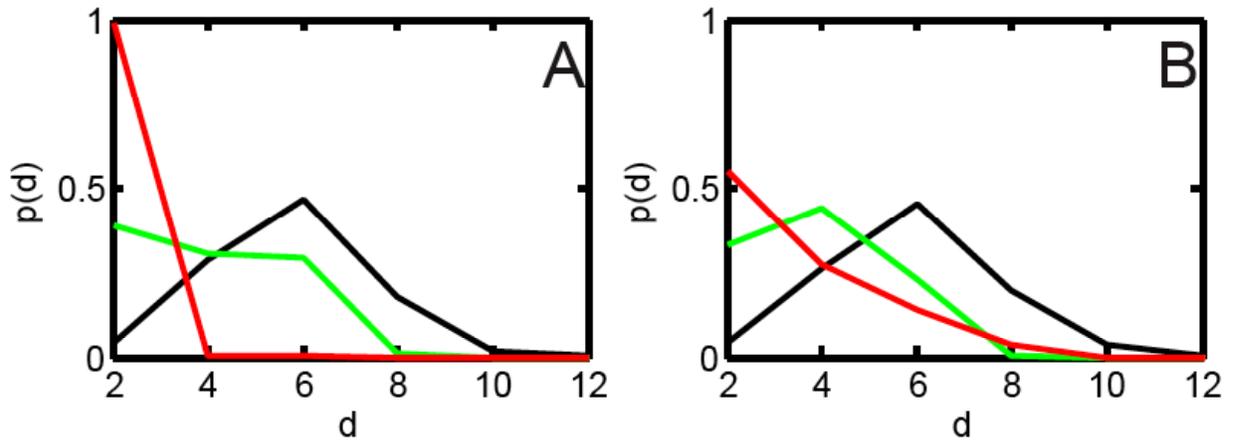

**Figure S6. Distance distribution of pairs of genes**

Distance distribution of pairs on a network evolved in (**A**) static and (**B**) quasi-static environment. Red: synthetic lethal pairs, green: knockdown suppressor pairs, black: random pairs.

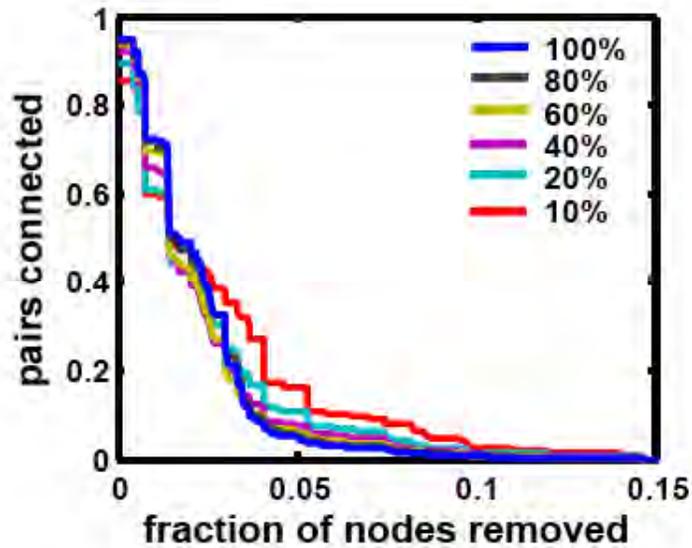

**Figure S7. Robustness of decay of knockdown suppressor pairs**

Fraction of knockdown suppressor pairs separated upon removing nodes with high BC using all (100%, weakest criterion) or fewer (only the top 10-80%) of suppressor pairs. The top 95% of pairs were used for Figures 8A and S5. See legend for colors and thresholds.

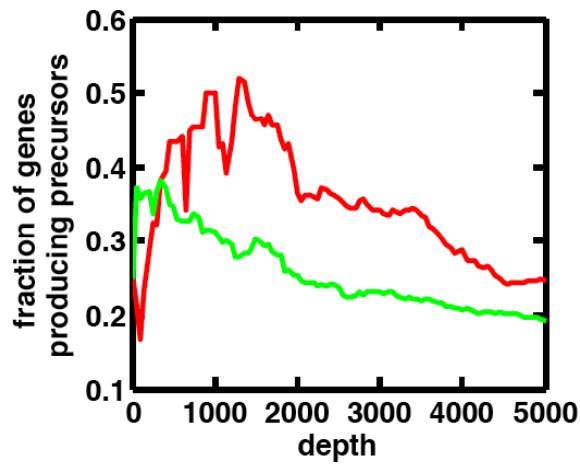

**Figure S8. Fraction of genes producing precursors**
Fraction of genes involved in the production of one of the 53 precursor molecules for the network evolved in a dynamic environment (red) vs. a static environment (green line).

## Supporting Tables
**Table S1**. **Organization of a 72 bp gene**

| Base pair | Parameter |
|---|---|
| 0-3 | Start |
| 4-7 | Expression level (converted to real number between 0 and 1) |
| 8-11 | Type of protein, obtained by taking the base 4 modulus of the sequence. 00=import, 01=export, 10=reaction, 11=reaction |
| 12-23 | Specificity of protein |
| 24-35 | Affinity domain 1 |
| 36-47 | Affinity domain 2 |
| 48-59 | Affinity domain 3 |
| 60-71 | Affinity domain 4 |